\documentstyle[11pt]{article}
\let\a=\alpha

\def\nn{\nonumber} \def\bd{\begin{document}} \def\ed{\end{document}}
\def\ds{\documentstyle} \let\fr=\frac \let\bl=\bigl \let\br=\bigr
\let\Br=\Bigr \let\Bl=\Bigl 
\let\bm=\bibitem
\let\na=\nabla
\let\pa=\partial \let\ov=\overline 
\newcommand{\be}{\begin{equation}} 
\newcommand{\ee}{\end{equation}} 
\def\ba{\begin{array}}
\def\ea{\end{array}}
\def\ft#1#2{{\textstyle{{\scriptstyle #1}\over {\scriptstyle #2}}}}
\def\fft#1#2{{#1 \over #2}}
\def\del{\partial}
\def\vp{\varphi}
\def\sst#1{{\scriptscriptstyle #1}}
\def\oneone{\rlap 1\mkern4mu{\rm l}}
\def\td{\tilde}
\def\wtd{\widetilde}
\def\ie{\rm i.e.\ }
\newcommand{\ho}[1]{$\, ^{#1}$}
\newcommand{\hoch}[1]{$\, ^{#1}$}
\newcommand{\bea}{\begin{eqnarray}} 
\newcommand{\eea}{\end{eqnarray}} 
\newcommand{\ra}{\rightarrow}
\newcommand{\lra}{\longrightarrow}
\newcommand{\Lra}{\Leftrightarrow}
\newcommand{\ap}{\alpha^\prime}
\newcommand{\bp}{\tilde \beta^\prime}
\newcommand{\tr}{{\rm tr} }
\newcommand{\Tr}{{\rm Tr} } 
\newcommand{\NP}{Nucl. Phys. }
\newcommand{\tamphys}{\it  Center for Theoretical Physics\\
Texas A\&M University, College Station, Texas 77843}
\newcommand{\ens}{\it Laboratoire de Physique Th\'eorique de l'\'Ecole
Normale Sup\'erieure\hoch{2}\\
24 Rue Lhomond - 75231 Paris CEDEX 05}
\newcommand{\auth}{N. Khviengia\hoch{\dagger},
Z. Khviengia\hoch{\dagger},  H. L\"u\hoch{\ddagger},
and C.N. Pope\hoch{\dagger1}}

\begin{document}
\begin{flushright}
\hfill{CTP TAMU-13/97}\\
\hfill{LPTENS-97/07}\\
\hfill{hep-th/9703012}\\
\hfill{March 1997}\\
\end{flushright}

\vspace{15pt}

\begin{center}
{ \large {\bf Towards a Field Theory of F-theory}}

\vspace{20pt}

\auth

\vspace{15pt}

{\hoch{\dagger}\tamphys}

\vspace{10pt}
{\hoch{\ddagger}\ens}

\vspace{40pt}

\underline{ABSTRACT}
\end{center}

     We make a proposal for a bosonic field theory in twelve dimensions that
admits the bosonic sector of eleven-dimensional supergravity as a consistent
truncation.  It can also be consistently truncated to a ten-dimensional
Lagrangian that contains all the BPS $p$-brane solitons of the type IIB
theory.  The mechanism  allowing the consistent truncation in the latter
case is unusual, in that  additional fields with an off-diagonal kinetic
term are non-vanishing and yet do not contribute to the dynamics of the
ten-dimensional theory.  They do, however, influence the oxidation of
solutions back to twelve dimensions.  We present a discussion of the
oxidations of all the basic BPS solitons of M-theory and the type IIB string
to $D=12$.  In particular, the NS-NS and R-R strings of the type IIB theory
arise as the wrappings of membranes in $D=12$ around one or other circle of
the compactifying 2-torus.

{\vfill\leftline{}\vfill
\vskip	10pt
\footnoterule
{\footnotesize
     \hoch{1}	Research supported in part by DOE 
Grant DE-FG03-95ER40917 \vskip	-12pt} \vskip 10pt
{\footnotesize
        \hoch{2} Unit\'e Propre du Centre National de la Recherche
Scientifique, associ\'ee \`a l'\'Ecole Normale Sup\'erieure 
\phantom{abcdef}et \`a
l'Universit\'e de Paris-Sud}
}

\pagebreak
\setcounter{page}{1}

\section{Introduction}
 
     For a variety of reasons, most notable of which is its $SL(2,Z)$ 
symmetry, it has been proposed that the type IIB string in ten dimensions 
may have its origin in a twelve-dimensional theory known as F-theory 
\cite{hull,vafa}.  This proposal is parallel to the suggestion that the type 
IIA string has its origin in eleven-dimensional M-theory \cite{wit}.  In this 
latter case, it is clear that the low-energy limit of M-theory is the 
long-known eleven-dimensional supergravity \cite{cjs}.  Indeed, one can see 
evidence for the eleven-dimensional origin of the type IIA string purely at 
the level of its low-energy effective action, in that this ten-dimensional
type  IIA supergravity arises as a consistent truncation that retains just
the massless sector of the compactification of $D=11$ supergravity on a
circle \cite{cw,hn}.

     In the case of the type IIB string and F-theory, we do not have the 
benefit of already knowing the form of the low-energy effective theory in 
twelve dimensions.  Indeed, we know that there cannot be any ordinary
supergravity field theory in $D=12$.  Thus we have a less clear-cut starting
point for investigations of the compactifications of the hypothetical
F-theory.  One may hope nevertheless that even at a purely
field-theoretic level, there might exist some twelve-dimensional theory that
could be compactified to ten dimensions on a 2-torus, in such a way that
type IIB supergravity would emerge as a consistent truncation.  It is
crucial that the truncation should be consistent, in order that solutions of
the type IIB supergravity will also be solutions of the equations of motion
of the twelve-dimensional theory in which it is embedded.  The 
twelve-dimensional theory should also be capable of being consistently 
truncated to $D=11$ supergravity.

     In this paper we shall study this problem from the field theoretic 
point of view, along the lines discussed above.  In particular, we shall 
present arguments that seem to lead rather naturally to a candidate for the 
twelve-dimensional field theory.  To be more precise, we focus our 
attention on the bosonic sector of the field theory.   We first show that it
can be consistently truncated to the bosonic sector of $D=11$ supergravity, 
after compactifying from $D=12$ on a circle.  We then consider instead the 
compactification from $D=12$ to $D=10$ on a 2-torus.  We show that this 
admits a consistent truncation to the set of fields considered in the 
Lagrangian formulation of the type IIB theory in \cite{bbo}, namely the 
fields of type IIB supergravity but with no duality restriction on the 
5-form field strength.  (In \cite{bbo}, the equations of motion following 
from the Lagrangian with the non-self-dual 5-form can be reduced to the
type IIB equations by imposing self-duality as a consistent truncation.)
Our dimensionally reduced theory in ten dimensions, after truncating the other 
unwanted fields, admits type IIB supergravity as a further consistent 
truncation at the linearised level.  To be more specific, all interaction 
terms except the trilinear interaction between the 5-form and the two 3-form 
field strengths are correctly reproduced.   Thus the subset of type IIB 
solutions for which this interaction plays no r\^ole are also solutions of 
our ten-dimensional equations of motion. A crucial 
ingredient in the construction turns out to be a dilatonic scalar field 
in the twelve-dimensional theory, which plays a vital r\^ole in the 
subsequent truncation in ten dimensions. 

    Given embeddings of $D=11$ supergravity and type IIB supergravity in the 
twelve-dimensional theory, one can consider the various BPS-saturated 
$p$-brane solutions in these lower-dimensional theories, and examine their 
reinterpretation as solutions in the higher dimension.  We are able to carry 
out this procedure to oxidise the M-branes of M-theory to $D=12$.  Although 
the incompleteness of the embedding of the type IIB theory precludes a 
full discussion of oxidations from $D=10$, one may nevertheless argue that 
the oxidation to $D=12$ of those solutions of type IIB gravity for which the
trilinear gauge-field interaction plays no r\^ole may be sensibly discussed.
These include all the basic BPS-saturated $p$-branes of the type IIB theory.
In particular, we show how the NS-NS and R-R strings of the type IIB theory
arise from wrapping membranes in $D=12$ around one or other circle of the
compactifying 2-torus. 

\section{The low-energy effective Lagrangian for F-theory}

     Several authors have discussed the probable field content of a
twelve-dimensional theory that could be capable of yielding type IIB
supergravity after compactification to ten dimensions
\cite{fms,tseyt,kar}.  As will be the case in our subsequent
considerations, attention has been restricted to the bosonic fields in
the theory.  It is clear that in order to obtain a 5-form field
strength in $D=10$, the simplest assumption would be to include such a
5-form $G_5$ already in $D=12$. For two reasons, this is not yet
sufficient, however.  Firstly, we need to get two 3-form field
strengths in $D=10$, whilst the toroidal reduction of $G_5$ would
yield only one.  Secondly, we need to obtain the right cubic
gauge-field interaction term in $D=11$.  This would have to come from
a topological cubic term in $D=12$, and there is no way to write such
a term if one has only $G_5$ and its potential $B_4$ available.  Both
of these deficiencies might be cured if a second field strength
$F_4=dA_3$, of rank 4, is introduced too.  Now we may write down a
topological term $B_4\wedge dA_3\wedge dA_3$ in $D=12$ \cite{fms},
which is capable of yielding cubic terms of the right general structures in
eleven and ten dimensions.  Thus our proposed bosonic field content in
$D=12$ now includes the metric $g_{\sst{MN}}$, and the 3-form and 4-form
potentials $A_3$ and $B_4$. 
 
     It is not hard to see that there is still a further deficiency with 
the above field content.  The reason for this can be seen by looking at the 
way in which Kaluza-Klein reductions work in general.  When a field theory 
is compactified on a circle from $D+1$ dimensions to $D$ dimensions, the 
metric in the Einstein frame is reduced according to
\be
ds_{\sst D+1}^2 = e^{2\a\varphi}\, ds_{\sst D}^2 + e^{-2(D-2)\a\varphi} \,
(dz +{\cal A})^2\ ,\label{metred}
\ee
where $ds_{\sst D}^2$, $\varphi$ and ${\cal A} ={\cal A}_{\sst M}\, dx^{\sst 
M}$ are independent of the compactification coordinate $z$, and the constant 
$\a$ is chosen to be $\a=((2(D-1)(D-2))^{-1/2}$ in order to achieve a 
conventional normalisation for the lower-dimensional kinetic terms.  A 
potential of rank $(n-1)$ is reduced according to
\be
A_{n-1}(x,z)\longrightarrow A_{n-1}(x) + A_{n-2}(x)\wedge dz\ .
\label{formred}
\ee
This means that if we start from a Lagrangian in $D+1$ dimensions of the 
form
\be
{\cal L}_{\sst D+1}= e\, R -\ft12e\, (\del\phi)^2 -\fft{1}{2\, n!}\, e \, 
e^{\hat a\phi}\, F_n^2\ ,\label{dp1lag}
\ee
then the resulting Lagrangian in $D$ dimensions will be \cite{lpss}
\bea
{\cal L}_{\sst D} &=& e\, R -\ft12 e\, (\del\phi)^2 -\ft12 e\,
(\del\varphi)^2 -\ft14 e\, e^{-2(D-1)\a\varphi}\, {\cal F}_2^2 \nn\\
&&-\fft{1}{2\, n!}\, e\, e^{\hat a\phi -2(n-1)\a\varphi}\, F_n^2 
-\fft{1}{2\, (n-1)!}\, e\, e^{\hat a\phi+2(D-n)\a\varphi}\, F_{n-1}^2 
\ ,\label{dplag}
\eea
where ${\cal F}_2 = d{\cal A}$. Note that the lower-dimensional field
strength $F_n$ appearing here will acquire a Chern-Simons type correction,
with $F_n=dA_{n-1} - dA_{n-2}\wedge {\cal A}$ in $D$ dimensions. This can
be seen by noting that the exterior derivative of (\ref{formred}),
organised in terms of vielbein components, is $dA_{n-1} -dA_{n-2}\wedge
{\cal A} + dA_{n-2}\wedge (dz + {\cal A})$.  It is these vielbein
components which appear in the lower-dimensional kinetic terms.

     The strength of a dilaton coupling of the form $e^{\hat a\phi +b\varphi}
F_n^2$ can best be expressed by making a rotation to a new 
canonically-normalised dilatonic scalar 
$\phi'= (\hat a^2+b^2)^{-1/2}\, (\hat a\phi + b\varphi)$, so that we now  
have $e^{a\phi'}\, F_n^2$, where $a^2=\hat a^2 + b^2$.  Although the value
of  the coupling constant $a$ is changed from the value $\hat a$ in the
higher  dimension, there is a nice way to reparameterise the couplings in a
way that  is preserved under Kaluza-Klein reduction \cite{lpss}.  Thus we
introduce a new constant $\Delta$, related to $a$ in $D$ dimensions by 
\be
a^2 =\Delta - \fft{2(n-1)(D-n-1)}{D-2}\ .\label{delta}
\ee
It is not hard to see that the effective dilaton couplings for all field 
strengths are such that their values of $\Delta$ remain unchanged under 
Kaluza-Klein reduction.  One can also easily verify that the new 2-form 
field strengths ${\cal F}_2$ that emerge from the dimensional reduction of
the metric have $\Delta=4$.  In fact, interestingly enough {\it all} dilaton 
coupling constants for all field strengths in maximal supergravities have 
$\Delta=4$ \cite{lpss}.  This includes cases such as $D=11$
supergravity,  where the absence of the dilaton corresponds to having
$a=0$,  and this translates, using (\ref{delta}), into $\Delta=4$.  

     With these preliminaries, we can see why the Kaluza-Klein reduction of 
a twelve-dimensional field theory involving just $g_{\sst{MN}}$, $A_3$ and 
$B_4$ is not capable of admitting type IIB supergravity as a consistent 
truncation.  The reason is that the type IIB theory has $\Delta=4$ for all 
its dilaton couplings, but the theory in $D=12$ with no dilaton will, using 
(\ref{delta}), have $\Delta=\ft{21}{5}$ for the 4-form field strength, and 
$\Delta=\fft{24}{5}$ for the 5-form field strength.  Unless some truncation 
of the field content in the dimensionally-reduced theory is performed, the 
values of $\Delta$ in the higher dimension will persist in the lower 
dimension.  Another possibility is to introduce a dilaton already in 
$D=12$, in order to change the couplings to $\Delta=4$ already in
twelve dimensions.  It turns out that this latter procedure is the one that 
is needed in order to end up with the correct $\Delta=4$ dilaton couplings 
in ten dimensions.  On the other hand, the required $\Delta=4$ coupling of 
$D=11$ supergravity can be achieved without the need for the dilaton in 
$D=12$, by setting an appropriate linear combination of the two 4-form field 
strengths in eleven dimensions to zero.  This truncation can also be 
described within the framework needed for the ten-dimensional reduction, by 
simply setting the $D=12$ dilaton to zero as well.  Thus we may embed both 
the eleven-dimensional and ten-dimensional theories in a single theory in 
$D=12$, where a dilaton is included in order to have couplings with 
$\Delta=4$.  Accordingly,  we take as our starting point the 
twelve-dimensional Lagrangian density
\be
{\cal L}_{12} = e\, R -\ft12\, e\, (\del\psi)^2 - \ft{1}{48}\, e^{a\psi}\,
F_4^2 -\ft{1}{240}\, e\, e^{b\psi}\, G_5^2 + \lambda\, B_4\wedge dA_3\wedge
dA_3\ ,
\label{d12lag}
\ee
where $\psi$ is the dilaton, and $a$ and $b$ are constants.  The last term 
is understood to be dualised to give a 0-form contribution to the 
Lagrangian density, {\it i.e.}\ it is equal to $\ft{\lambda}{864}\,
\epsilon^{\sst M_1\cdots \sst M_{12}}\, B_{\sst M_1\cdots \sst M_4}\,
\del_{\sst M_5} A_{\sst M_6\cdots \sst M_8} \del_{\sst M_9}A_{\sst
M_{10}  \cdots \sst M_{12}}$.

   Using (\ref{delta}), it is easy to see that if the dilaton couplings to 
the 4-form and 5-form field strengths are to correspond to the
canonical value $\Delta=4$, we must have
\be
a^2=-\ft15\, \qquad\qquad b^2=-\ft45\ .\label{abval}
\ee
It is interesting that to achieve $\Delta=4$, it seems that theories in
$D<11$ need real dilaton couplings, the theory in $D=11$ itself needs zero
dilaton coupling, and the theories in $D>11$ need imaginary dilaton
couplings.  We shall see later that the imaginary couplings, far from being
undesirable, are exactly what is needed in order to make a consistent
truncation to the fields of type IIB supergravity possible. 

    Let us now follow the standard Kaluza-Klein procedure to reduce the 
twelve-dimensional theory (\ref{d12lag}), first to $D=11$ and then to $D=10$.
In an obvious notation, where $F_4$ reduces to $F_4$, $F_3^{(i)}$, 
$F_2^{(ij)},\ldots$ after compactification on internal circles labelled by 
$i$, and $G_5$ similarly reduces to $G_5$, $G_4^{(i)}$, $G_3^{(ij)},\ldots$, 
we see from (\ref{dp1lag}) and (\ref{dplag}) that the reduction of
(\ref{d12lag}) to $D=11$ gives 
\bea
e^{-1}{\cal L}_{11} &=& R -\ft12(\del\psi)^2 -\ft12(\del\phi_1)^2 -\ft1{48}\,
e^{-\fft{1}{\sqrt5}\phi_1 + a\psi}\, F_4^2 \nn\\
&&-\ft1{12} e^{\fft{7}{3\sqrt5}\phi_1 +a\psi}\, (F_3^{(1)})^2 
-\ft1{240}\, e^{-\fft{4}{3\sqrt5} \phi_1 + b\psi}\, G_5^2 \label{d11lag}\\
&&-\ft1{48}\, e^{\fft{2}{\sqrt5} \phi_1 +b\psi}\, (G_4^{(1)})^2 
-\ft14\, e^{-\fft{10}{3\sqrt5}\phi_1}\, ({\cal F}_2^{(1)})^2 +
e^{-1}{\cal L}_{\rm FFA}\ .\nn
\eea
As mentioned above, there will be Chern-Simons type modifications in the
expressions for the field strengths $F_4$ and $G_5$ in eleven dimensions.
Their detailed forms may be found by applying the results given in 
\cite{lpsol}.  We shall discuss the cubic interactions ${\cal L}_{\rm FFA}$
later.

     After a reduction on a further circle, we will get a Lagrangian in ten 
dimensions which now has three dilatonic scalars, namely $\psi$, $\phi_1$ 
and a further scalar $\phi_2$ from the $g_{11,11}$ component of the 
eleven-dimensional metric.  It will be convenient to perform an $SO(2)$ 
rotation on the $\phi_1$ and $\phi_2$ dilatons, by defining
\be
\phi=\fft{\sqrt5}{3}\, \phi_1 -\fft23\, \phi_2\ ,\qquad\qquad
\varphi=\fft23\, \phi_1 +\fft{\sqrt5}{3}\, \phi_2 \ .\label{phirot}
\ee
As we shall see below, it is the $\phi$ field that acquires an
interpretation as the dilaton of the type IIB theory.  The $\varphi$ field, 
on the other hand, parameterises the volume of the compactifying 2-torus,
which is given by $e^{-\ft{2}{\sqrt5}\varphi}$.
In terms of these rotated scalars, we find that the dimensional reduction of 
(\ref{d12lag}) to $D=10$ gives the Lagrangian
\bea
e^{-1}{\cal L}_{10} \!\!\!&=&\!\!\! R -\ft12(\del\psi)^2
-\ft12(\del\phi)^2 -\ft12 
(\del\varphi)^2 -\ft12 e^{-2\phi}\, (\del\chi)^2 -\ft1{12}
e^{\fft{3}{\sqrt5}\varphi + b\psi}\, (G_3^{(12)})^2 \nn\\ 
\!\!\! && \!\!\! -\ft1{240}\, e^{-\fft{2}{\sqrt5}\varphi +b\psi}\,
G_5^2  -\ft1{12}
e^{\phi+\fft1{\sqrt5}\varphi + a\psi}\, (F_3^{(1)})^2 -
\ft1{12} e^{-\phi +\fft1{\sqrt5}\varphi +a\psi}\, (F_3^{(2)})^2 \nn\\
\!\!\! && \!\!\! -\ft1{48}e^{-\fft3{2\sqrt5}\varphi+a\psi}\, F_4^2 
-\ft1{48}e^{\phi+\fft1{2\sqrt5}\varphi+b\psi}\, (G_4^{(1)})^2 
-\ft1{48} e^{-\phi+\fft1{2\sqrt5}\varphi+b\psi}\, 
(G_4^{(2)})^2\label{d10lag}\\
\!\!\! && \!\!\! -\ft14 e^{\fft7{2\sqrt5}\varphi+a\psi}\, (F_2^{(12)})^2 
-\ft14 e^{-\phi-\fft{\sqrt5}{2}\varphi}\, ({\cal F}_2^{(1)})^2 
-\ft14 e^{\phi-\fft{\sqrt5}{2}\varphi}\, ({\cal F}_2^{(2)})^2 
+e^{-1}{\cal L}_{\rm FFA}\ ,\nn
\eea
where $\chi={\cal A}_0^{(12)}$ is the 0-form potential coming from the 
dimensional reduction of the Kaluza-Klein vector ${\cal A}_1^{(1)}$ in 
$D=11$.  There will be Chern-Simons modifications in several of the field
strengths appearing here; again, the full details may be found using the
general results in \cite{lpsol}.

      Having obtained the complete dimensionally-reduced Lagrangians in
$D=11$ and $D=10$ dimensions, we shall now examine the possibility of
consistently truncating them to give $D=11$ supergravity and type IIB
supergravity respectively.  First, we shall consider the truncation to
$D=11$ supergravity.  This can be performed exactly (at least in the
bosonic sector, which we are considering in this paper).  In this
truncation, the dilatonic scalar $\psi$ which we introduced in twelve
dimensions is in fact set to zero.  In the truncation of the
ten-dimensional Lagrangian (\ref{d10lag}), on the other hand, it
will turn out that the dilaton $\psi$ plays a crucial r\^ole.

\subsection{Reduction to $D=11$ supergravity}

     In this section we shall show that the dimensional reduction of the
twelve-dimensional theory to $D=11$ may be truncated to give $D=11$
supergravity.  Our starting-point is the eleven-dimensional Lagrangian
(\ref{d11lag}).  Obviously, since (\ref{d11lag}) contains more fields
than are present in $D=11$ supergravity, some of them must be set to
zero.  The important point is that this truncation must be {\it
consistent}, \ie setting the fields to zero must be consistent with
their equations of motion.  We first observe that we may consistently
set
\be
G_5=F_3^{(1)}= {\cal F}_2^{(1)} =0\ .\label{truncate1}
\ee
In general, owing to the Chern-Simons modifications and ${\cal L}_{\rm
FFA}$ term, setting a field strength to zero can be inconsistent with
the equations of motion.   However in this case, setting all three of the
field strengths in (\ref{truncate1}) to zero simultaneously is nevertheless
consistent.  This follows from the fact that their equations of motion
have the general form
\bea
\nabla \cdot G_5 &\sim& \epsilon F_4 F_3^{(1)}\ ,\nonumber\\
\nabla \cdot F_3^{(1)} &\sim& \epsilon G_5 F_4 + F_4 \cdot {\cal
F}_2^{(1)} \ ,\label{eomtrun}\\
\nabla \cdot {\cal F}_2^{(1)} &\sim& G_5 \cdot G_4^{(1)} +
F_4 \cdot F_3^{(1)}\ ,\nonumber
\eea
and thus after imposing (\ref{truncate1}), the source terms on the
right-hand sides of the equations of motion (\ref{eomtrun}) vanish.

     Now recall that we are choosing the dilaton coupling constants
$a$ and $b$ to satisfy (\ref{abval}), which we solve by taking 
\be
a=\fft{i}{\sqrt5}\, \qquad\qquad b=-\, \fft{2i}{\sqrt5} \ .\label{abvalue}
\ee
Let us define the complex field $w= (-\phi_1 + i\psi)/\sqrt2$, $\bar w=
(-\phi_1-i\psi)/\sqrt2$, in terms of which the consistently truncated
Lagrangian takes the form
\be
e^{-1}{\cal L}_{11} = R -\del w\cdot\del \bar w - \ft{1}{48}
e^{\sqrt{\fft25} w}\, F_4^2 -\ft1{48} e^{-2\sqrt{\fft25} w}\, (G_4^{(1)})^2
+ \lambda\,e^{-1}\,  B_3^{(1)}\wedge dA_3\wedge dA_3\ .\label{d11lag2}
\ee
Note that the anti-holomorphic part $\bar w$ 
of the complex scalar field appears only in the off-diagonal kinetic term,
and thus the equations of motion for $w$ and $\bar w$ are
\bea
\nabla^2 w &=& 0\ ,\nn\\
\nabla^2\bar w &=& \ft1{48}\, \sqrt{\ft25}\, \Big(e^{\sqrt{\fft25} w}\, 
F_4^2 -2 e^{-2 \sqrt{\fft25} w}\, (G_4^{(1)})^2 \Big)\ .\label{wbweom}
\eea
It should be emphasised that $w$ and $\bar w$ are to be treated as
independent fields.\footnote{Another way of describing the theory is by 
starting out with a dilatonic scalar in $D=12$ with the ``wrong sign'' for its
kinetic term.  Effectively we may define $\theta=i\psi$, so that in $D=12$ we
have 
$$
{\cal L}_{12} = e\, R +\ft12\, e\, (\del\theta)^2 - \ft{1}{48}\,e\,
e^{\fft{1}{\sqrt5}\theta}\, F_4^2 -\ft{1}{240}\, e\,
e^{-\fft{2}{\sqrt5}\theta}\, G_5^2 +\lambda B_4\wedge dA_3\wedge dA_3\ .    
$$
Now, the combinations of $\phi_1$ and $\theta$ in $D=11$ will be of the
form $w=(-\phi_1+\theta)/\sqrt2$ and $\tilde w=(-\phi_1-\theta)/\sqrt2$, again
with the desired off-diagonal kinetic term $-\del w\cdot \del \tilde w$ in the
eleven-dimensional Lagrangian.  In this formulation it is clear that $w$ and
$\tilde w$ are independent, and that we may set $w=0$ while $\tilde w$ is
non-zero.  Since the complex description is formally equivalent to this one,
we shall for simplicity work with the $w$ and $\bar w$ variables.}  Clearly,
we may consistently set $w=0$, and while (\ref{wbweom}) implies that $\bar w$
will in general be  non-zero, it is a kind of ``phantom'' field that does
not influence any of  the other eleven-dimensional equations of motion.

     From (\ref{d11lag2}) the equations of motion for $F_4$ and $G_4^{(1)}$,
having set $w=0$, become
\bea
d*F_4 &=& 2\lambda\, G_4^{(1)}\wedge F_4\ ,\nn\\
d*G_4^{(1)} &=& \lambda\, F_4\wedge F_4\ .\label{4feom}
\eea
In order to obtain $D=11$ supergravity, it is necessary to reduce the
remaining system of fields further, so that in particular we have only a
single independent 4-form field strength, rather than two. We may do this by
taking $F_4$ and $G_4^{(1)}$ to be proportional, again making sure that this
truncation of the theory is consistent with the equations of motion.  Thus we
may define
\be
F_4 = \wtd F_4 \, \sin\beta\, \qquad\qquad G_4 = \wtd F_4\, \cos\beta\ ,
\label{f4trun}
\ee
where the constant angle $\beta$ is determined by the consistency of the two
equations (\ref{4feom}), which then imply $F_4=\sqrt2\, G_4^{(1)}$ and hence
\be
\cos\beta = \fft{1}{\sqrt3}\ ,\qquad\qquad \sin\beta = \sqrt{\fft23}\ .
\ee
Substituting into (\ref{wbweom}), we find that the right-hand side of the
equation of motion for $\bar w$ in fact now vanishes, and thus we may solve
$\nabla^2 \bar w=0$ by taking $\bar w=0$.   By this means we arrive at a
consistent truncation of the dimensionally reduced eleven-dimensional theory
described by (\ref{d11lag}),  which is described by the Lagrangian
\be
{\cal L} = e R - \ft{1}{48}\,e\,  \wtd F_4^2 + \ft2{3\sqrt3}\, \lambda\,
\wtd A_3\wedge d\wtd A_3\wedge d\wtd A_3\ ,
\ee
where $\wtd F_4 = d\wtd A_3$.  Comparison with the Lagrangian for the bosonic
sector of $D=11$ supergravity shows that we must take the coefficient
$\lambda$, introduced in (\ref{d12lag}), to be
\be
\lambda = \ft14\, \sqrt 3\ .
\ee
Thus we have succeeded in embedding the bosonic sector of $D=11$ supergravity
as a consistent truncation in the twelve-dimensional theory  described by
(\ref{d12lag}).

   It is worth emphasising that the dilaton $\psi$ in twelve 
dimensions played an essential r\^ole in ensuring the consistency of the 
truncation of the fields given in (\ref{truncate1}), since the complex field 
$\bar w=(-\phi_1-i\psi_1)/\sqrt2$ had to be non-zero in order to satisfy its 
equation of motion (\ref{wbweom}).  The fact that the further truncation 
(\ref{f4trun}) eventually led to the simpler situation where $\psi$ vanished 
was a consequence of the apparent coincidence that the ratio between $F_4$
and $G_4^{(1)}$ that was required for consistency of the equations of motion
for these two fields happened to be the same as the ratio for which the
right-hand side of the equation for $\bar w$ in (\ref{wbweom}) vanished. 
The situation is different, however, in the truncation of the fields in the
ten-dimensional theory (\ref{d10lag}) to those of type IIB supergravity.  In
the next section, we show that in this case the consistency of the 
truncation to the desired set of fields requires that $\psi$ be non-zero. 

\subsection{Reduction to type IIB supergravity}

     In this section we shall attempt the truncation of the
ten-dimensional Lagrangian (\ref{d10lag}) to give type IIB
supergravity.  We begin by setting to zero those field strengths 
that are not present in type IIB supergravity, namely
\be
F_4=G_4^{(1)}=G_4^{(2)} = G_3^{(12)} = {\cal F}_2^{(1)} =
{\cal F}_2^{(2)} =F_2^{(12)} =0\ ,\label{zero1}
\ee
As in the eleven-dimensional case, we find from a careful
investigation of the equations of motion for these fields that their
non-linear source terms, arising because of the Chern-Simon
modifications and ${\cal L}_{\rm FFA}$ term, vanish when the
conditions (\ref{zero1}) are imposed, and thus the truncation of these
fields is a consistent one.  The Lagrangian (\ref{d10lag}) is now
reduced to 
\bea
e^{-1}{\cal L}_{10} &=& R -\ft12(\del\psi)^2 -\ft12(\del\phi)^2 -\ft12 
(\del\varphi)^2 -\ft12 e^{-2\phi}\, (\del\chi)^2
-\ft1{240}\, e^{-\fft{2}{\sqrt5}\varphi +b\psi}\, G_5^2\nonumber\\
&& -\ft1{12} e^{\phi+\fft1{\sqrt5}\varphi + a\psi}\, (F_3^{(1)})^2 -
\ft1{12} e^{-\phi +\fft1{\sqrt5}\varphi +a\psi}\, (F_3^{(2)})^2
+e^{-1}{\cal L}_{\rm FFA}\ .\label{d10lag2}
\eea
After taking into account the Chern-Simons modifications to the various
field strengths, as discussed above, we find that $F_3^{(1)}$ and
$F_3^{(2)}$ are given by
\be
F_3^{(1)} = dA_2^{(1)}\ ,\qquad\qquad F_3^{(2)} = dA_2^{(2)} - \chi\,
dA_2^{(1)} \ .\label{f3cs}
\ee
These are precisely the structures of the NS-NS and R-R 3-forms
respectively, in type IIB supergravity.  Note that before the truncations 
(\ref{zero1}) there are in total three 3-form field strengths, namely 
$G_3^{(12)}$, $F_3^{(1}$ and $F_3^{(2)}$.  Since $G_3^{(12)}$ is a singlet 
under the $SL(2,R)$ symmetry, it is clear that it should be truncated from 
the ten-dimensional theory; the remaining two 3-forms form the required 
doublet under $SL(2,R)$.

     We would like to be able to truncate the scalars $\varphi$ and $\psi$
also, since only the dilaton $\phi$ should remain in the type IIB theory.  
However, it is clear from (\ref{d10lag2}) that setting $\varphi$ and $\psi$ 
to zero would be inconsistent with their equations of motion, since the 
remaining field strengths act as sources for these scalars.  At this 
point, we recall that the constants $a$ and $b$ should satisfy the 
conditions given in (\ref{abval}).  In particular, we shall take them to be
given by (\ref{abvalue}).  If we then define the complex field $u$, and its
conjugate $\bar u$, by
\be
u=\fft1{\sqrt2} (\varphi + i\psi)\ ,\qquad\qquad \bar u=
\fft1{\sqrt2} (\varphi-i\psi)\ ,\label{udef}
\ee
then the truncated Lagrangian (\ref{d10lag2}) takes the form
\bea
e^{-1}{\cal L}_{10} &=& R -\del u\cdot \del \bar u -\ft12(\del\phi)^2  
 -\ft12 e^{-2\phi}\, (\del\chi)^2
-\ft1{240}\, e^{-2\sqrt{\fft{2}{5}}\, u} \, G_5^2 \nn\\
&&-\ft1{12}e^{\sqrt{\fft{2}{5}}\, u} \Big( e^{\phi}\, (F_3^{(1)})^2 +
e^{-\phi}\, (F_3^{(2)})^2 \Big)
+e^{-1}{\cal L}_{\rm FFA}\ .\label{d10lag3}
\eea
The equations of motion for $u$ and $\bar u$ are
\bea
\nabla^2 u &=& 0\ ,\label{ueq}\\
\nabla^2 \bar u &=& \ft1{12}\sqrt{\ft{2}{5}}\, e^{\sqrt{\fft{2}{5}}\, u}\,
\Big( e^{\phi}\, (F_3^{(1)})^2 + e^{-\phi}\, (F_3^{(2)})^2 \Big)
-\ft1{120} \sqrt{\ft{2}{5}}\, e^{-2\sqrt{\fft{2}{5}}\, u}\, G_5^2
\ .\label{ubeq}
\eea
Note that $u$ and $\bar u$ are treated as independent variables in these
equations (just like the $w$ and $\bar w$ fields in $D=11$), and they are
both $SL(2,R)$ invariant.  We see that it is consistent to set $u=0$, while
$\bar u$ will in general be non-zero. However, the important point is that
$\bar u$ appears in the Lagrangian only in the off-diagonal kinetic term
$\del u\cdot\del \bar u$, and so $\bar u$ is a phantom field which, even
though non-zero,  has no influence on the solutions for the other fields in
the ten-dimensional theory.  

     In terms of $\varphi$ and $\psi$, we see from (\ref{udef}) that the
volume $e^{-\ft2{\sqrt5}\varphi}$ of the compactifying 2-torus is not 
constant in general.  However, it does not contribute to the dynamics of the 
ten-dimensional theory, and it does not interfere with the $SL(2,R)$ 
symmetry.  It should again be emphasised that although it might {\it a priori}
have seemed more natural to demand that $\varphi$ simply be non-dynamical, 
it is not possible to do this because its equation of motion requires that 
it be non-constant in general.  Of course if one considers only the scalar 
sector of the type IIB theory, where the higher-degree field strengths are 
zero, then the inconsistency of setting $\varphi=$const. in the full theory 
is not apparent.  Thus the necessity of introducing the dilaton $\psi$ in 
$D=12$, which resolves this inconsistency, is not seen if one restricts 
attention to the scalar sector.  For this reason the discussion in 
\cite{tseyt}, where the 3-form and 5-form field strengths were taken to be 
zero, does not encounter inconsistencies.  However the theory in $D=12$ with 
the field content $\{g_{\sst{MN}}, A_3,B_4\}$ proposed in \cite{fms,tseyt,kar} 
will run into this inconsistency problem.  Note that the inconsistency would 
arise not only in the $T^2$ compactification of the twelve-dimensional 
theory, but also in all the compactifications on elliptically-fibred 
manifolds.  It should be emphasised also that despite the fact that $u$ and
$\bar u$ are treated as independent variables, the volume parameter
$\varphi$ will always be real, and it is only the dilaton $\psi$ that is
imaginary.  As we remarked earlier, this also could be made real by replacing
$\psi$ by a field $\theta=i\psi$. 

         In order to obtain the precise form of the ten-dimensional
type IIB supergravity theory, we should like to be able to perform a
further truncation, by setting the anti-self-dual part of the 5-form
field strength $G_5$ to zero.  In fact the Lagrangian (\ref{d10lag3}),
after taking care of the truncation of $u$ and the decoupling of $\bar
u$ discussed above, is superficially of the form of the Lagrangian
obtained in \cite{bbo}, which, when properly used by imposing
self-duality of $G_5$ after obtaining the equations of motion, describes
type IIB supergravity.  However, there is unfortunately a discrepancy,
namely that our field strength $G_5$ in $D=10$ is given simply by
$G_5= dB_4$, whereas the field strength in \cite{bbo} has a
Chern-Simons correction, with $G_5 = dB_4 + \lambda\, \epsilon_{ij} A_2^{(i)}
\wedge dA_2^{(j)}$.  This means that since our equation of motion and
Bianchi identity for $G_5$ are of the form 
\be
d*G_5 =\lambda \, \epsilon_{ij} dA_2^{(i)} \wedge dA_2^{(j)}\ , 
\qquad dG_5 = 0\ ,\label{eombi}
\ee
we cannot in general, unlike in \cite{bbo} where $dG_5=\lambda \, 
\epsilon_{ij} dA_2^{(i)} \wedge dA_2^{(j)}$, consistently
impose the self-duality condition $G_5=*G_5$.

       It should be remarked that we would not necessarily have to be able to 
set the anti-self-dual 5-form to zero, provided that it decoupled from the 
other equations of motion.  Let us define $H_5^{(\pm)} = G_5 \pm *G_5$,
where $H_5^{(+)}$ is self-dual and $H_5^{(-)}$ is anti-self-dual.    Then the
self-dual 5-form $H_5^{(+)}$ satisfies precisely the same equation of motion
and Bianchi identity as in type IIB supergravity, namely \cite{sch2}
\be
d*H_5^{(+)}=dH_5^{(+)} =\lambda\, \epsilon_{ij} dA_2^{(i)} \wedge dA_2^{(j)}\ .
\label{sd5f}
\ee
Unfortunately, the anti-self-dual 5-form $H_5^{(-)}$ does not decouple
from the full equations of motion of type IIB supergravity; it appears in
the equations of motion for the 3-form field strengths, namely
\be
d*F_3^{(i)} =2\lambda\,  G_5 \wedge F_3^{(j)} \epsilon_{ij} =       
 \lambda\, (H_5^{(+)} + H_5^{(-)}) \wedge F_3^{(j)} \epsilon_{ij}
\ .
\ee

     The truncation of the twelve-dimensional Lagrangian to the type
IIB theory in D=10 is therefore consistent only up to linear order
when the 5-form field strength is involved.  However, the problem is
avoided altogether in the case of configurations for which
\be
\epsilon_{ij} dA_2^{(i)} \wedge dA_2^{(j)} = 0\ .\label{ffterm}
\ee
For precisely the same reason, it is possible to consider a ``truncated''
Lagrangian for type IIB supergravity without the 5-form field strength
\cite{sch}.  In general such a truncation is inconsistent, owing to the
equations of motion and Bianchi identity for the self-dual 5-form field
strength $H_5^{(+)}$, given in (\ref{sd5f}); however, it becomes consistent
if (\ref{ffterm}) is satisfied. 

    This condition can easily be seen to be satisfied by all singly-charged
BPS-saturated $p$-brane solitons, and their $SL(2,R)$ duality multiplets, in
the type IIB theory, and so such solutions of the type IIB theory (which
preserve half the supersymmetry) are also solutions of the
dimensionally-reduced Lagrangian (\ref{d10lag3}) with $u=0$.   In type IIB
supergravity, there is, first of all, a self-dual 3-brane, which makes use
of the R-R self-dual 5-form field strength.  The solution is a singlet under
the $SL(2,R)$ symmetry, and all the other field strengths and the dilaton
$\phi$ are zero.  Thus we may consistently impose the self-dual constraint
in our dimensionally-reduced Lagrangian in this case.  For the remaining
single-charge $p$-branes and their $SL(2,R)$ multiplets, using 3-form or
1-form field strengths, the constraint (\ref{ffterm}) is again satisfied and
the 5-form field strength is zero.  It follows that the corresponding
truncated equations of motion are precisely the same as those of type IIB
supergravity.  Thus all the BPS-saturated extremal $p$-branes of the type
IIB theory are contained in the dimensionally-reduced Lagrangian coming from
$D=12$.  Further analysis shows that in fact {\it all} BPS solutions in all 
dimensional reductions of the type IIB theory also satisfy (\ref{ffterm}) in 
$D=10$, and thus they are also all solutions of our ten-dimensional theory.
These include multiply-charged solutions that preserve smaller fractions of 
the supersymmetry, as well as the singly-charged ones that preserve half the 
supersymmetry.

    Finally, we remark that we could also take the full ten-dimensional theory 
(\ref{d10lag}), and perform an alternative consistent truncation that would
give the bosonic sector of type IIA supergravity.  Of course this would be
nothing but a restatement in $D=10$ of the truncations that we performed in
$D=11$ to get eleven-dimensional supergravity.

\section{Oxidation of M-branes to $D=12$}

      In this section, we consider the BPS-saturated membrane and 5-brane
solutions of the low-energy effective limit of M-theory, and examine
their oxidations to the twelve-dimensional field theory presented in
the previous section.  It is, of course, crucial that we have been
able to embed $D=11$ supergravity as a {\it consistent} truncation of
the theory in $D=12$, in order that this notion of oxidation should
have a well-defined meaning.

         We begin with the  electrically-charged membrane in $D=11$, 
which takes the form \cite{dust}
\bea
ds_{11}^2 &=& H^{-2/3}\, (-dt^2 + dx^i\, dx^i) +H^{1/3}\, dy^m\, dy^m\ ,
\nn\\
F_{m\mu\nu\rho} &=& \epsilon_{\mu\nu\rho}\, \del_m H^{-1}\ ,\label{memsol}
\eea
where $i=1,2$ and $H$ is harmonic in the 8-dimensional transverse space of
the $y^m$ coordinates.  The simplest single-membrane solution has $H=1+k\,
r^{-6}$, where $r^2=y^m\, y^m$.  As we saw in the previous section, the
consistent truncation of (\ref{d11lag}) to $D=11$ supergravity implies that
$w=\bar w=0$, and hence $\psi=\phi_1=0$.  Using (\ref{metred}), we therefore 
find that the oxidation of the  eleven-dimensional membrane gives simply
\be
ds_{12}^2 = H^{-2/3}\, (-dt^2 + dx^i\, dx^i) + H^{1/3}\, 
dy^m\, dy^m  +  dz_1^2\label{memox}
\ee
Since the additional term $dz_1^2$ does not have the harmonic functional 
dependence either of the membrane world-volume or of the transverse space, 
it describes neither a 3-brane nor a line of membranes in $D=12$. 

   If we instead begin from the 5-brane solution in M-theory, we have \cite{gu}
\bea
ds_{11}^2 &=& H^{-1/3}\, (-dt^2 + dx^i\, dx^i) +H^{2/3}\, dy^m\, dy^m\ ,
\nn\\
F_{mnpq} &=& \epsilon_{mnpqr}\, \del_r H\ ,\label{5bsol}
\eea
where $i=1,\ldots,5$ and $H$ is harmonic in the 5-dimensional transverse
space.  The oxidation to $D=12$ is simply given by
\be
ds_{12}^2 = H^{-1/3}\, (-dt^2 + dx^i\, dx^i) + H^{2/3}\, 
dy^m\, dy^m + dz_1^2\ . \label{5box}
\ee
Again, this describes neither a 6-brane nor a line of 5-branes in $D=12$.

         Note that the M-branes supported by the field strength
$\widetilde F_4$ in fact carry charges associated both with $F_4$ and
$G_4^{(1)}$.  However, owing to the absence of a dilaton in $D=11$, the
M-branes cannot be viewed as bound states of these two charges with zero
binding energy, unlike the case of the self-dual 3-brane in $D=10$ which we
shall discuss in section 4.  Thus the twelve-dimensional interpretation of
M-branes is somewhat obscured.  The membrane in $D=11$, when wrapped around 
the 11'th coordinate, gives rise to the NS-NS
string in type IIA theory, and has a quite different interpretation in
$D=12$ from the NS-NS string in type IIB, which, as we shall discuss in
section 4, arises as a line of membranes in $D=12$, wrapped around the 12'th
coordinate $z_1$.  This may not be surprising since after all the NS-NS
string in type IIA is intrinsically different from the NS-NS string in type
IIB in $D=10$.  They do, however, become equivalent owing to T-duality upon
compactification to $D=9$. 

\section{Oxidation of type IIB $p$-branes to $D=12$}

     To begin, we shall consider the string solution in $D=10$, supported by the
NS-NS 3-form field strength $F_3^{(1)}$.  The equations of motion for the
metric, $\phi$ and $F_3^{(1)}$ are the standard ones for the type IIB
string, giving the usual string solution
\bea
ds_{10}^2&=& H^{-3/4}\, (-dt^2 + dx^2) + H^{1/4}\, 
dy^m dy^m \ ,\nn\\ 
e^\phi&=& H^{1/2}\ ,\qquad \qquad F_{m\mu\nu}= \epsilon_{\mu\nu}\,
\del_m H^{-1}\ ,\label{stringnsns}
\eea
where $H$ is an harmonic function in the 8-dimensional transverse
space described by the $y^m$ coordinates. For an isotropic string, the 
harmonic function is given by 
$H=1 + k r^{-6}$.  In order to oxidise the solution (\ref{stringnsns}) back
to $D=12$, we need to obtain the solution for $\bar u$.  The equation of motion
(\ref{ubeq}) for $\bar u$ becomes 
\be
\nabla^2 \bar u = \ft1{12} \sqrt{\ft{2}{5}}\, e^\phi\, (F_3^{(1)})^2\ ,
\ee
which, after substituting the form of the solution (\ref{stringnsns}), is
readily seen to imply that $\bar u = \fft1{\sqrt{10}}\, \log H$.  Since
the truncation to the type IIB theory sets $u=0$, it follows from
(\ref{udef}) that we have $e^{2\sqrt5\varphi}= H$.  Thus from
(\ref{phirot}) and (\ref{stringnsns}), we find that the Kaluza-Klein scalars
$\phi_1$ and $\phi_2$ coming from the reduction on the 2-torus are given by
\be
e^{\phi_1} = H^{\fft{7}{6\sqrt5}}\ ,\qquad\qquad 
e^{\phi_2} = H^{-1/6}\ .
\ee
Finally, tracing back through the steps of dimensional reduction of the
metric using (\ref{metred}), we arrive at the twelve-dimensional oxidation
of the type IIB NS-NS string metric
\be
ds_{12}^2 = H^{-7/10} (-dt^2 + dx^2 + dz_1^2) + H^{3/10}\, (dy^m dy^m
+ dz_2^2).\label{stringoxnsns}
\ee
Here, for simplicity, we have taken the two additional dimensions to be
spacelike.  At least in some versions of F-theory, it would be more
natural to perform a Wick rotation on one of them, to give a second
timelike coordinate (similar remarks apply to the additional coordinate $z_1$
in the oxidations of M-theory discussed in the previous section).  
Note that the metric (\ref{stringoxnsns})
describes a twelve-dimensional membrane (\ie an F-brane) with world volume
coordinates $(t, x, z_1)$, and with charges uniformly distributed along the
transverse-space coordinate $z_2$.   This result is understandable, since the
NS-NS 3-form field strength $F_3^{(1)}$ comes from the dimensional reduction
of the 4-form field strength $F_4$ in $D=12$, which admits a membrane
solution. To get the string in $D=10$ from a membrane in $D=12$, one step of
vertical dimensional reduction and one step of diagonal dimensional
reduction are necessary. 

          Now let us see how the ten-dimensional R-R string solution
oxidises to $D=12$.  In this case, the solution is supported by
the R-R 3-form field strength $F_3^{(2)}$.  The metric and the field
strength in the solution in $D=10$ are the same as those given in
(\ref{stringnsns}), but the dilaton is given by $e^{-\phi} = H^{1/2}$.  
Since now the equation of motion for  $\bar u$ is $\nabla^2 \bar u =
\ft1{12} \sqrt{\fft{2}{5}}\, e^{-\phi}\,  (F_3^{(2)})^2$, it follows that
the solution for $\varphi$ is identical to the NS-NS case, namely
$e^{2\sqrt5 \varphi} = H$. This is not surprising since as we observed in
section 2, both $u$ and $\bar u$ are $SL(2,R)$ invariant.  From these
expressions for $\phi$ and $\varphi$, one can easily convert to the
$(\phi_1, \phi_2)$ basis using (\ref{phirot}).  Thus the twelve-dimensional
metric for the R-R string solution becomes 
\be
ds_{12}^2 = H^{-7/10} (-dt^2 + dx^2 + dz_2^2) + H^{3/10}\, (dy^m dy^m
+ dz_1^2).\label{stringoxrr}
\ee
This describes a line of membranes uniformly distributed along the $z_1$
direction.  It is interesting to compare the two twelve-dimensional metrics
(\ref{stringoxnsns}) and (\ref{stringoxrr}).  When the twelve-dimensional
membrane wraps around the 12'th coordinate $z_1$, it gives rise to an NS-NS
string in the ten-dimensional type IIB theory, whereas when the membrane
wraps instead around the 11'th coordinate $z_2$, it gives rise to an R-R
string.  This is consistent with the proposition that the non-perturbative
$SL(2,Z)$ symmetry \cite{sch} of the type IIB theory, which rotates between
the NS-NS and R-R strings, is the symmetry of the 2-torus on which the
F-theory is compactified. 

    We have seen that the NS-NS and R-R strings of the type IIB theory arise 
as twelve-dimensional membranes wrapped around different circles in the 
2-torus.  Thus the $SL(2,Z)$ symmetry of the 2-torus is reflected naturally 
in the $SL(2,Z)$ symmetry that relates the NS-NS and R-R strings.  The
membrane in $D=12$ is supported by the electric charge carried by the 4-form 
field strength $F_4$.  The two string solutions, corresponding to the 
wrapping of the membrane around either $z_1$ or $z_2$, are understood from 
the ten-dimensional point of view as carrying charge under either 
$F_3^{(1)}$ or $F_3^{(2)}$.  There is also a
3-brane solution in $D=12$, with electric charge carried by the 5-form $G_5$. 
(This will have a (2,2) world-volume signature if the signature of the
spacetime is (10,2).) This 3-brane can be wrapped around the entire 2-torus to
also give a string in $D=10$.  However in this case it is a singlet under
$(SL(2,R)$, and hence is not part of the type IIB spectrum.  Indeed, it
corresponds to a string supported by the singlet 3-form field strength
$G_3^{(12)}$ that we truncated from the dimensionally-reduced theory in $D=10$
in order to retain only the fields of the type IIB theory.\footnote{Note
that this wrapping of the 3-brane in $D=12$ around the entire 2-torus does
not correspond to the type IIA string either.  As we saw in the previous
sections, the 4-form field strength of $D=11$ supergravity is a linear
combination of $G_4^{(1)}$ and $F_4$, and thus neither the membrane in $D=11$
nor its diagonal reduction to the string in type IIA has its origin in any
simple F-brane solution in $D=12$.  Instead, the solution is given by
(\ref{memox}).} Thus the $SL(2,R)$ structure of the type IIB strings seems to
be explained naturally in terms of a membrane origin in
$D=12$, rather than the 3-brane origin discussed in \cite{vafa}.  The situation
is precisely analogous to that of the $SL(2,R)$ doublet of string solitons in
maximal nine-dimensional supergravity; one carries an NS-NS charge, whilst the
other carries an R-R charge.  They are obtained from M-theory by wrapping the
membrane around one or other of the circles on the compactifying 2-torus.  

     It is interesting to look at the more general family of $(p,q)$ string 
solutions in the type IIB theory \cite{sch}, where the NS-NS and R-R 3-forms 
carry charges $p$ and $q$ respectively.  These solutions can be obtained by 
performing an $SL(2,R)$ rotation of the pure $(1,0)$ NS-NS string.  The 
ten-dimensional metric is invariant under this transformation, and is the 
same as the one given in (\ref{stringnsns}).  The dilaton $\phi$ and axion 
$\chi$ are now given by
\be
e^{-\phi}= a^2\, H^{-1/2} + b^2\, H^{1/2}\ , \qquad\qquad
\chi=\fft{ac+ bd\, H}{a^2 + b^2\, H}\ ,\label{sl2r}
\ee
where $ad-bc=1$.  The solution for $\bar u$, being $SL(2,R)$ invariant, is 
unchanged by this transformation.  After oxidising back to $D=12$, we obtain 
the metric
\bea
ds_{12}^2 &=& H^{-7/10}\, (-dt^2+ dx^2) + H^{3/10}\, dy^m\, dy^m +
(a^2 + b^2 \, H)^{-1} \, H^{3/10}\, dz_2^2 \ ,\nn\\
&&+(a^2 + b^2\, H) H^{-7/10}\, \Big(dz_1 + \fft{ac + bd\, H}{ a^2 + b^2\, H} 
\, dz_2\Big)^2 \ .\label{sl2rox}
\eea

         An analogous analysis applies to the NS-NS and R-R 5-branes. 
They are given by
\be
ds_{10}^2 = H^{-1/4} (-dt^2 + dx^i dx^i) + H^{3/4} dy^m dy^m\ ,\quad
F_{mnp} = \epsilon_{mnpr}\del_r H\ ,
\ee
together with $e^{-\phi} = H^{1/2}$ for the
NS-NS 5-brane and $e^{\phi} = H^{1/2}$ for the R-R 5-brane.  Here
$H$ is an harmonic function on the four-dimensional transverse space
of the coordinates $y^m$.   It is straightforward, using
(\ref{ueq}), (\ref{ubeq}) and (\ref{udef}), to show that
$e^{-2\sqrt5\varphi}=H$.  Tracing back through the steps of
dimensional reduction, we find that the metrics for the NS-NS and R-R
5-branes in $D=12$ are given by
\bea
ds_{12}^2 &=& H^{-3/10} (-dt^2 + dx^i dx^i + dz_1^2) + H^{7/10}\, (dy^m dy^m
+ dz_2^2)\ , \label{5braneoxnsns}\\
ds_{12}^2 &=& H^{-3/10} (-dt^2 + dx^i dx^i + dz_2^2) + H^{7/10}\, (dy^m dy^m
+ dz_1^2)\ ,\label{5braneoxrr}
\eea
respectively.   As one would expect from our previous discussion for strings, 
both the NS-NS and R-R 5-branes in
$D=12$ are dimensional reductions of a line of 6-branes in $D=12$,
supported by a magnetic charge for the 4-form field strength $F_4$.  When
the 6-brane wraps around $z_1$, it gives rise to the NS-NS 5-brane in the
type IIB theory; when it wraps instead around $z_2$, the resulting solution
is the R-R 5-brane.    

              There are three more $p$-branes in the $D=10$ type
IIB theory, namely the self-dual 3-brane using the self-dual 5-form field
strength \cite{hs,dl}, and the instanton and 7-brane using the 1-form field 
strength $\del \chi$ \cite{ggp}. For all of these solutions, it
follows  from (\ref{ueq}) and (\ref{ubeq}) that we have $u=0 = \bar u$.
Let us first consider the the self-dual 3-brane, in which case
all the dilatons $(\phi_1, \phi_2, \psi)$ are zero, and so the
twelve-dimensional metric is given by 
\be
ds_{12}^2 = ds_{10}^2 + dz_1^2 + dz_2^2\ , 
\ee
where $dz_{10}^2$ is the ten-dimensional metric of the self-dual
3-brane. This solution is obviously consistent with the fact that the
self-dual 3-brane is invariant under the non-perturbative $SL(2,Z)$
symmetry of the type IIB string.  Naively, one might expect that the
self-dual 3-brane could be viewed as a bound state of electric and
magnetic 3-branes, which could be oxidised to a 3-brane and a 5-brane in
$D=12$ respectively, and hence that the self-dual 3-brane could be
viewed as an intersection of a 3-brane and a 5-brane in $D=12$.
However, as was observed in \cite{kklp}, the self-dual 3-brane in type
IIB theory has $\Delta=4$, and is therefore a basic state itself which cannot
be viewed as a bound state of the $\Delta=4$ electric and magnetic
3-branes with zero binding energy.  In fact, it was shown in
\cite{lpsol} that the metric of a dyonic 3-brane in a non-self-dual
theory in $D=10$ is given by
\be
ds_{10}^2 = (1 + k r^{-6})^{-1/2} (-dt^2 + dx^i dx^i) +
            (1 + k r^{-6})^{1/2} (dr^2 + r^2 d\Omega_5)\ ,
\ee
where $k=\sqrt{Q_e^2 + Q_m^2}$, with $Q_e$ and $Q_m$ the electric and
magnetic charges.   Thus the self-dual 3-brane can instead be viewed  as a
bound state with positive binding energy, which cannot be interpreted as an
intersection of $p$-branes in higher dimensions \cite{lptax}. 

        Finally, we have the instanton solution, which oxidises to a
pp-wave in $D=12$ \cite{tseyt},  and the 7-brane
solution.  In this latter case, if a 24-centre configuration is chosen, 
the $D=12$ metric contains a K3 metric, which can be viewed as a 2-torus
bundle over a 2-sphere \cite{vafa}.

\section{Discussion and Conclusions}

           In this paper, we have shown that there exists a possible
candidate field theory for F-theory in $D=12$, whose
dimensional reduction to $D=11$ admits a consistent truncation to the
bosonic sector of eleven-dimensional supergravity.  In particular, this
allows the M-branes of M-theory to be oxidised back to solutions in F-theory.
The same twelve-dimensional theory also admits an inequivalent
truncation, after dimensional reduction to $D=10$, that coincides in
many respects with the bosonic sector of type IIB supergravity.
Specifically, solutions of type IIB supergravity for which the
bilinear quantity $\epsilon_{ij} dA_2^{(i)} \wedge dA_2^{(j)}$
vanishes will also be solutions of the dimensionally-reduced
theory in $D=10$.  One may hope that in configurations where the
constraint vanishes, which can thus simultaneously be solutions of
type IIB supergravity and the twelve-dimensional theory, the embedding
again gives a valid oxidation of the lower-dimensional solutions to
those of F-theory.  The consistency of the truncation of fields in the 
ten dimensional theory requires the presence of a dilaton already in $D=12$, 
which, together with a linear combination $\varphi$ of the Kaluza-Klein 
dilatons $\phi_1$ and $\phi_2$, forms a complex scalar field $u$ that plays a 
rather unusual r\^ole.  Specifically, the truncation allows $u$ to be 
set to zero, while $\bar u$ is required to be non-vanishing, although  it is
non-dynamical.  It does, however, participate in the oxidation of solutions 
back to $D=12$.

    The main defect of the embedding scheme that we have presented in this
paper is that the theory in $D=10$ to which the compactified twelve-dimensional
theory can be truncated is not quite the same as type IIB supergravity.  
However, its solution set has a considerable overlap with the solutions of
the type IIB theory, including all of the BPS-saturated $p$-brane
solitons in $D\le10$.  One may hope that this lacuna can be overcome in a way
that does not spoil this already successful embedding of the BPS solitons.
This will be the case provided that the complete embedding differs from the
one that we have presented here by terms that vanish when 
$\epsilon_{ij} dA_2^{(i)} \wedge dA_2^{(j)}$ is zero.  Possible ideas for 
obtaining an exact embedding of the type IIB theory include introducing a
Chern-Simons type modification of the form $G_5 = dB_4 +\kappa\, *(A_3\wedge
dA_3)$ in $D=12$, which would lead to a modified 5-form field strength of
the form $G_5=dB_4 +\kappa\,  *(\epsilon_{ij} A_2^{(i)} \wedge dA_2^{(j)})$
in $D=10$. Another rather similar idea is to include a 7-form field strength
already in $D=12$ (again with a dilaton coupling with $\Delta=4$), which
could take the form $G_7=dB_6 + \kappa\, A_3\wedge dA_4$.  In fact, neither
of these possibilities seems to give the desired result in $D=10$, although
they both show some promising features.  In any case, provided that a
complete description of the type IIB embedding can be found by some
modification along these lines, the discussion of the oxidation of type IIB
$p$-branes to $D=12$ in the previous section should continue to be valid.  A
further issue that we have not tackled in this paper is the inclusion of a
fermionic sector in the twelve-dimensional theory.  It may well be that this
would single out a preferred signature for the additional 12'th coordinate,
which in our present discussion could be either spacelike or timelike. 

     Another unexplained aspect of the reduction procedures that we have 
described in this paper concerns the truncations of fields that we needed to 
perform in $D=11$ and $D=10$.  Although we do not have a completely 
satisfactory explanation for why they should be performed, it is worth 
emphasising that there are rather tight constraints on what sets of fields 
can be consistently truncated from the dimensionally reduced theories.  As 
we saw in $D=11$, the interaction terms in the Lagrangian imply that entire 
sets of fields must be truncated simultaneously, and so the number of 
possible truncations is severely limited.  

    A further observation that puts the truncations on a sounder basis is 
that there is in fact a symmetry principle that selects the fields 
that are set to zero in (\ref{truncate1}) and (\ref{zero1}).  Let us
consider the eleven-dimensional case first.  It is easy to see that after
imposing the conditions (\ref{abvalue}) on the dilaton couplings in $D=12$,
then substituting the definitions of $w$ and $\bar w$ into the full
eleven-dimensional Lagrangian (\ref{d11lag}), we shall find that the fields
listed in (\ref{truncate1}) will be precisely the ones whose exponential
prefactors in (\ref{d11lag}) will include $\bar w$ dependence.  It follows
that the Lagrangian (\ref{d11lag}) has a global scaling symmetry in which we
send $\bar w \rightarrow \bar w+$const. together with appropriate
non-trivial scalings of all the fields listed in the truncation
(\ref{truncate1}).  On the other hand the other fields that we are retaining
in the Lagrangian (\ref{d11lag2}) are invariant under this symmetry.  Now it
is always the case that if one truncates the fields in a Lagrangian to a
subset comprising all the singlets under a symmetry group, then the
truncation will be consistent \cite{dupo}.  This is because the singlet fields
that are retained cannot act as sources for the non-singlets that are
truncated.  Similarly, in $D=10$ the set of fields that are truncated in
(\ref{zero1}) are precisely the subset that scale non-trivially under a
global symmetry where $\bar u\rightarrow\bar u +$const. in the full $D=10$
Lagrangian (\ref{d10lag}).  The fact that there is a symmetry principle in
both $D=11$ and $D=10$ that selects the truncated fields implies that one
may hope that the consistency of these truncations will also persist beyond
the level of the classical field theory. 

     We have described a twelve-dimensional theory that contains more
bosonic degrees of freedom than are seen in its M-theory or type IIB theory 
dimensional reductions, thus necessitating the imposition of
consistent truncations in $D=11$ or $D=10$.  Of course one could take an
alternative viewpoint, and impose precisely these field truncations already
in $D=12$.  Needless to say, they cannot be imposed in a twelve-dimensionally
covariant manner, and so in this kind of a formulation the field theory in
$D=12$ would have only eleven-dimensional or ten-dimensional covariance.  It
may therefore be merely a matter of convenience as to whether one prefers to
work with a covariant twelve-dimensional theory with extra degress of freedom,
or non-covariant theories with the correct degrees of freedom.  After
including the fermionic sector, one may similarly have such a choice of
descriptions.  In a twelve-dimensionally covariant form, the theory would,
for example, reduce to a non-supersymmetric theory in $D=10$ that admitted a
consistent truncation to $N=2$ type IIB supergravity.  Alternatively, by
sacrificing twelve-dimensional covariance, one might be able to construct a
supergravity theory that was already supersymmetric in $D=12$.  However,
since the truncation needed to obtain eleven-dimensional supergravity is
inequivalent to the truncation needed to obtain type IIB supergravity, it
would seem that the two would only be unified in $D=12$ by taking
the covariant twelve-dimensional theory as the starting point.

\section*{Acknowledgment}

     We are grateful to K. Benakli and M.J. Duff for useful discussions.

\vfill\eject

\end{document}